# Segmentation analysis on a multivariate time series of the foreign exchange rates

Aki-Hiro SATO[1, a]

[1]Department of Applied Mathematics and Physics, Graduate School of Informatics,
Kyoto University, Yoshida Honcho, Sakyo-ku, 606-8501 Kyoto JAPAN

[a]sato.akihiro.5m@kyoto-u.ac.jp

**Abstract.** This study considers the multivariate segmentation procedure under the assumption of the multivariate Gaussian mixture. Jensen-Shannon divergence between two multivariate Gaussian distributions is employed as a discriminator and a recursive segmentation procedure is proposed. The daily log-return time series for 30 currency pairs consisting of 12 currencies for the last decade (January 3, 2001 to December 30, 2011) are analyzed using the proposed method. The proposed method can detect several important periods related to the significant affairs of the international economy.

## Introduction

Over the last two decades, the statistical properties of asset price returns have been successively studied in the literature of econophysics [1, 2]. One important property is that the probability distribution of returns exhibits a fat-tailed distribution [3, 4]. In this study, I hope to provide some insights on the problem of finding transition points in the global economy. In the context of economics and finance, there are various methods that can be used to segment highly nonstationary financial time series into stationary segments called regimes or trends. Following the pioneering works of Goldfeld and Quandt [5], there is much literature on detecting structural breaks or change points separating stationary segments. Recently, a recursive entropic scheme to separate financial time series was proposed [6].

The multivariate time series can be modeled by using multivariate Gaussian distribution. However, in the case of financial time series, we normally observe the multivariate time series as a mixture of multivariate Gaussian distributions with a different variance-covariance matrix and mean due to the nonstationarity of variance-covariance matrix and mean values. Therefore, we obtain the variance-covariance matrix of an unconditional distribution when we compute an empirical variance-covariance matrix from all the data. Furthermore, the mixture of multivariate Gaussian distributions is normally a non-Gaussian distribution.

In this study, we consider a segmentation procedure for multivariate time series under the assumption of local stationarity. We assume that the multivariate time series are generated from different multivariate Gaussian distributions. The proposed procedure is applied to segmenting multiple daily log-return time series for 30 selected currency pairs from the period of January 4, 2001 to December 30, 2011. This article is organized as follows. In Sec. 2, the recursive segmentation procedure is briefly explained. In Sec. 3 the proposed method is applicable to segmenting a mixture of multivariate Gaussian samples with the given variance-covariance matrix. In Sec. 4, an empirical analysis with daily log-returns for the last 10 years is conducted. Sec. 5 is devoted to conclusions.

## Segmentation procedure

Let $r_i(t)$ ($i = 1,\ldots, M; t = 1,\ldots,T$) be $M$-dimensional multiple log-return time series, defined as $r_i(t) = R_i(t+1) - R_i(t)$, where $R_i(t)$ ($t = 1,\ldots,T+1$) is the daily exchange rate of $i$-th currency pair at day $t$. From successive works on financial markets, the log-return time series for foreign exchange rates are modeled by $q$-Gaussian distributions [2] and/or Lévy distributions [1,3]. Both the $q$-Gaussian and Lévy distributions are given by an infinite mixture of Gaussian distribution with Gamma distribution multipliers[2]. In the context of finance, the $q$-Gaussian distributions are referred as to Student-$t$ distributions.

Let us assume that this multivariate time series consist of $n$ sequences sampled from $n$ different multivariate Gaussian distributions. I further assume that the log-return movements in segment $k$ follow multivariate Gaussian distributions with variance-covariance matrix $\mathbf{C}^{(k)}$. To determine the $n$ stationary segments from the given multiple time series $r_i(t)$, I employ the recursive segmentation procedure introduced by Cheong et al. [6]. In this segmentation procedure, we check whether the likelihood value is suitable in order to separate the multiple time series into two segments at the point $t$. To do so, we denote the likelihood as

$$\Delta(t) = \log L_2(t) - \log L_1, \quad (1)$$

where

$$L_1 = \prod_{s=1}^{T} p(r_1(s),\ldots,r_M(s);\boldsymbol{\mu},\mathbf{C}),\ L_2(t) = \prod_{s=1}^{t} p(r_1(s),\ldots,r_M(s);\boldsymbol{\mu}^{(L)},\mathbf{C}^{(L)}) \prod_{s=t+1}^{T} p(r_1(s),\ldots,r_M(s);\boldsymbol{\mu}^{(R)},\mathbf{C}^{(R)}),$$

$$p(r_1,\ldots,r_M;\boldsymbol{\mu},\mathbf{C}) = \frac{1}{(2\pi)^{M/2}|\mathbf{C}|^{1/2}} \exp\left[-\frac{1}{2}\sum_{i=1}^{M}\sum_{j=1}^{M}[\mathbf{C}^{-1}]_{ij}(r_i(s)-\mu_i)(r_j(s)-\mu_j)\right].$$

Eq. (1) is approximated as

$$\Delta(t) = -\sum_{s=1}^{T}\log p(r_1(s),\ldots,r_M(s);\boldsymbol{\mu},\mathbf{C}) + \sum_{s=1}^{t}\log p(r_1(s),\ldots,r_M(s);\boldsymbol{\mu}_L,\mathbf{C}_L) + \sum_{s+t+1}^{T}\log p(r_1(s),\ldots,r_M(s);\boldsymbol{\mu}_L,\mathbf{C}_R)$$

$$\approx \frac{T}{2}\log|\mathbf{C}| - \frac{t}{2}\log|\mathbf{C}^{(L)}| - \frac{T-t}{2}\log|\mathbf{C}^{(R)}|.$$

By replacing the Gaussian parameters $\{\boldsymbol{\mu},\boldsymbol{\mu}^{(L)},\boldsymbol{\mu}^{(R)},\mathbf{C},\mathbf{C}^{(L)},\mathbf{C}^{(R)}\}$ with their maximum likelihood estimators $\{\hat{\boldsymbol{\mu}},\hat{\boldsymbol{\mu}}^{(L)},\hat{\boldsymbol{\mu}}^{(R)},\hat{\mathbf{C}},\hat{\mathbf{C}}^{(L)},\hat{\mathbf{C}}^{(R)}\}$, we can compute $\Delta(t)$ as

$$\Delta(t) \approx \frac{T}{2}\log|\hat{\mathbf{C}}| - \frac{t}{2}\log|\hat{\mathbf{C}}^{(L)}| - \frac{T-t}{2}\log|\hat{\mathbf{C}}^{(R)}|, \quad (2)$$

where

$$\begin{cases} \hat{\mu}_i = \frac{1}{T}\sum_{s=1}^{T}r_i(s),\ \hat{\mu}_i^{(L)} = \frac{1}{t}\sum_{s=1}^{t}r_i(s),\ \mu_i^{(R)} = \frac{1}{T-t}\sum_{s=t}^{T-t}r_i(s) \\ \hat{C}_{ij} = \frac{1}{T}\sum_{s=1}^{T}(r_i(s)-\mu_i)(r_j(s)-\mu_j),\ \hat{C}_{ij}^{(L)} = \frac{1}{t}\sum_{s=1}^{t}(r_i(s)-\mu_i)(r_j(s)-\mu_j),\ \hat{C}_{ij}^{(R)} = \frac{1}{T-t}\sum_{s=t}^{T-t}(r_i(s)-\mu_i)(r_j(s)-\mu_j) \end{cases}$$

$$(i = 1,\ldots,M; j = 1,\ldots,M)$$

$$(3)$$

If we compute $\Delta(t)$ at all possible times $t$, we find a spectrum of the Jensen-Shannon divergence. This spectrum has the maxima at $t^*$. Namely,

$$\Delta^* = \Delta(t^*) = \max_t \Delta(t). \quad (4)$$

Therefore, this point $t^*$ is adequate to separate the multiple time series into two statistically distinct segments. Each segment can be separated into two segments by using Eq. (3). At each stage of the segmentation, we can recursively separate the multiple time series into two segments. This is done iteratively until all segment boundaries have converged onto their optimal segment. We must therefore find the termination condition of this recursive procedure. Several termination conditions have been proposed in previous studies. In this study, I use the method of determining segment boundaries if their $\Delta(t)$ is larger than the amplitudes of typical fluctuations in the spectra based on Ref. [6]. To simplify the recursive segmentation, we adopted a conservative threshold of $\Delta_0=10M$, so that the recursive segmentation procedure is terminated if max $\Delta(t)$ is less than $\Delta_0$. More generally, by using $M$-dimensional probability densities $p, p_L,$ and $p_R$, we define

$$L_1 = \prod_{s=1}^{T} p(r_1(s),\ldots,r_M(s)), \quad L_2(t) = \prod_{s=1}^{t} p_L(r_1(s),\ldots,r_M(s)) \prod_{s=t+1}^{T} p_R(r_1(s),\ldots,r_M(s)),$$

and Eq. (1) can be also described as

$$\Delta(t) = -\sum_{s=1}^{T} \log p(r_1(s),\ldots,r_M(s)) + \sum_{s=1}^{t} \log p_L(r_1(s),\ldots,r_M(s)) + \sum_{s=t+1}^{T} \log p_R(r_1(s),\ldots,r_M(s))$$
$$\approx -TH[p] + tH[p_L] + (T-t)H[p_R],$$

where $H[p]$ is defined as Shannon entropy defined as

$$H[p] = -\int_{-\infty}^{\infty} dr_1 \cdots \int_{-\infty}^{\infty} dr_M \log p(r_1(s),\ldots,r_M(s)) p(r_1(s),\ldots,r_M(s)).$$

Since one has $p(r_1,\ldots,r_M) = \frac{t}{T} p_L(r_1,\ldots,r_M) + \frac{T-t}{T} p_R(r_1,\ldots,r_M)$ in the sense of empirical distribution, $\Delta(t)$ is proportional to the Jensen-Shannon divergence as

$$\Delta(t)/T = J_{\pi_1,\pi_2}(p_L, p_R) = H[\pi_L p_L + \pi_R p_R] - \pi_L H[p_L] - \pi_R H[p_R],$$

where the weight $\pi_1$ and $\pi_2$ are given as $\pi_1 = \pi_L = t/T$ and $\pi_2 = \pi_R = (T-t)/T$, respectively. Therefore, when $\Delta(t)$ is maximum, $p_L$ is the most different from $p_R$.

Furthermore, we mention an approximation error in terms of $t$. In order to compute $\Delta(t)$, we use the definition of empirical variance-covariance. Eq. (1) can be also described as

$$\Delta(t)/T \approx \frac{1}{2}\sum_{i=1}^{M} \log \lambda_i - \frac{\pi_L}{2}\sum_{i=1}^{M} \log \lambda_i^{(L)} - \frac{\pi_R}{2}\sum_{i=1}^{M} \log \lambda_i^{(R)}$$
$$\approx \frac{M}{2}\left( \int_0^{\infty} \rho(\lambda) \log \lambda d\lambda - \pi_L \int_0^{\infty} \rho^{(L)}(\lambda^{(L)}) \log \lambda^{(L)} d\lambda^{(L)} - \pi_R \int_0^{\infty} \rho(\lambda^{(R)}) \log \lambda^{(R)} d\lambda^{(R)} \right),$$

where $\lambda, \lambda_L,$ and $\lambda_R$ represent eigenvalues of variance-covariance matrix $\mathbf{C}, \mathbf{C}^{(L)}$, and $\mathbf{C}^{(R)}$, respectively. According to the relationship between the true covariance matrix and the empirical covariance matrix [8], if data matrix $r_i(t)$ are multivariate Gaussian samples and $T < 3M$, then independent of the actual covariance matrix of multivariate Gaussian variables, the eigenvalue density is approximated as the Marčenko-Pastur density,

$$\rho(\lambda) = \frac{T}{M}\frac{\sqrt{(\lambda_- - \lambda)(\lambda_+ - \lambda)}}{2\pi\sigma^2 \lambda} \qquad (\lambda_- \leq \lambda \leq \lambda_+), \qquad (5)$$

where $\lambda_\pm = \sigma^2\left(1 \pm \sqrt{\frac{M}{T}}\right)^2$ and $\sigma^2$ denotes a scale factor. Specifically, at $M=T=1$, Eq. (5) is described as

$$\rho(\lambda) = \frac{1}{2\pi\sigma^2}\sqrt{\frac{\lambda + \lambda_+}{\lambda}} \qquad (0 \leq \lambda \leq \lambda_+),$$

where $\lambda_+ = \sigma^2(1+\sqrt{2})^2$. As a result, $\Delta(t)$ includes a significant approximation error in this case. In the case of $M/T>1$, the density has a peak at $\lambda = 0$. Therefore, we have to find the adequate point $t^*$ restricting $t$, ranging from $t_{min}$ and $t_{max}$. The restriction is adjustable regarding a statistical significance of empirical variance-covariance matrix for $M/N$.

**Empirical analysis and discussion**

In the empirical analysis, I use daily log-returns of exchange rates for 30 ($M=30$) currency pairs[1] consisting of AUD, BRL, CAD, CHF, EUR, GBP, JPY, MXN, NZD, SGD, USD, and ZAR during the period of January 4, 2001, to December 30, 2011. There are 2,760 data points in the multiple time series. The proposed segmentation procedure is applied to separating the multiple log-return time series. We have 24 segments[2] at $\Delta_0=10M=300$. The restriction of searching is set as $t_{\min} = 3M+1=91$ and $t_{\max} = T\text{-}3M\text{-}1=T\text{-}91$. Furthermore, by using Eq. (3), we define entropy of the market at segment $k$ as

$$H = \tfrac{1}{2}\log\left((2\pi e)^M \left|\hat{\mathbf{C}}^{(k)}\right|\right),$$

where $\hat{\mathbf{C}}^{(k)}$ represents the empirical variance-covariance matrix computed from the multiple time series included in the segment $k$.

Paribas shock happened in segment 15. The Bear Sterns shock occurred in segment 16. The Lehman shock was noticeable in segment 17. Moreover, the Euro shock happened in segment 21. Fig. 1 (a) exhibits the entropy value at each segment. In segments 15 to 19, the entropy maintained was larger than other segments. This means that the uncertainty decreased (cross-correlations increased) due to the synchrony of the log-return time series over the foreign exchange market driven by the global financial crisis. After the global financial crisis, the uncertainty increased (cross-correlations decreases) again. The uncertainty also increased from segment 20. These periods correspond to the noticeable Euro shocks. We also show the first and second eigenvalues at each segment in Fig. 1 (b) and (c). Both the first and second eigenvalues are sensitive to the critical events. During the global financial crisis (from the Paribas shock to Lehman shock) and the Euro shock, the first eigenvalue was larger than during the other periods. The second eigenvalue was also larger both before and after the global financial crisis, but was large from 2001 till 2004.

The proposed method has an adjustable parameter $\Delta_0$. The segments obtained by this procedure depend on the value of $\Delta_0$. I confirmed the robustness of the segments obtained from different $\Delta_0$ values.

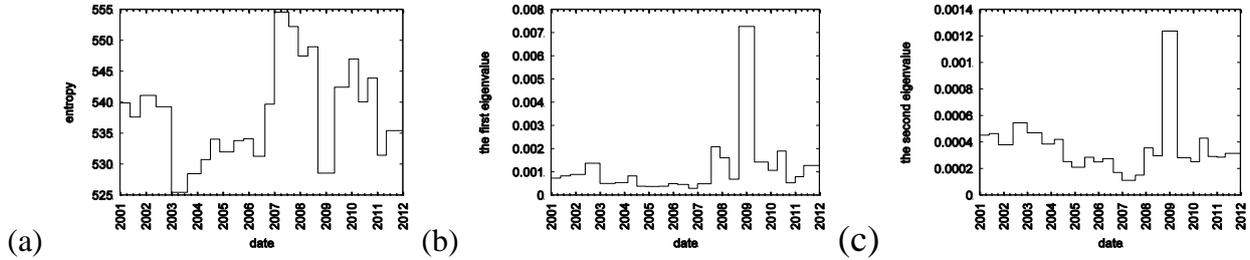

(a)  (b)  (c)

Fig.1. (a) The entropy of the 30 selected currency pairs of the foreign exchange market of the period of January 4, 2001, to December 30, 2011. (b) The first (b) and second (c) eigenvalues of the 30 selected currency pairs of the foreign exchange market of January 4, 2001, to December 30, 2011.

---

[1] The listed currency pairs are AUD/JPY, BRL/JPY, CAD/JPY, CHF/JPY, EUR/AUD, EUR/BRL, EUR/CAD, EUR/CHF, EUR/GBP, EUR/JPY, EUR/MXN, EUR/NZD, EUR/SGD, EUR/USD, EUR/ZAR, GBP/JPY, MXN/JPY, NZD/JPY, SGD/JPY, USD/AUD, USD/BRL, USD/CAD, USD/CHF, USD/GBP, USD/JPY, USD/MXN, USD/NZD, USD/SGD, USD/ZAR, and ZAR/JPY.

[2] I obtained Segment 1: 2001-01-03 to 2011-05-23, Segment 2:2001-05-24 to 2001-10-15, Segment 3:2001-10-16 to 2002-05-24, Segment 4: 2002-05-27 to 2002-12-31, Segment 5:2003-01-02 to 2003-08-15, Segment 6: 2003-08-18 to 2004-02-24, Segment 7:2004-02-25 to 2004-07-07, Segment 8: 2004-07-08 to 2004-11-17, Segment 9: 2004-11-18 to 2005-06-03, Segment 10: 2005-06-06 to 2005-10-24, Segment 11: 2005-10-25 to 2006-03-09, Segment 12:2006-03-10 to 2006-08-18, Segment 13:2006-08-21 to 2007-01-02, Segment 14:2007-01-03 to 2007-07-26, Segment 15:2007-07-27 to 2007-12-06, Segment 16:2007-12-07 to 2008-04-23, Segment 17:2008-04-24 to 2008-09-11, Segment 18: 2008-09-12 to 2009-05-04, Segment 19: 2009-05-05 to 2009-11-25, Segment 20: 2009-11-26 to 2010-04-09, Segment 21:2010-04-12to 2010-08-19,Segment 22: 2010-08-20 to 2010-12-31, Segment 23:2011-01-04 to 2011-05-11, and Segment 24:2011-05-12 to 2011-12-30.

**Summary**

I proposed a method to segment multivariate time series under the assumption of finite mixtures of multiple Gaussian distributions. Using numerical simulation, I confirmed that the proposed method was workable for a mixture of multivariate Gaussian samples. We conducted empirical analysis of daily log-return time series of 30 selected currency pairs of 4 January 4, 2011, to December, 30 2011. 24 segments were obtained by using the proposed method. I computed entropy and the first and second eigenvalues for each segment.

The proposed method could provide us time-dependent covariance of finite a multivariate Gaussian mixture. The first and second eigenvalues of the variance-covariance matrix of each regime showed the situation of the foreign exchange market. Such properties may be applied to detecting change points of the foreign exchange market.

**Acknowledgement**

This work is financially supported by the Grant-in-Aid for Young Scientists (B) by the Japan Society for the Promotion of Science (JSPS) KAKENHI (#23760074). The author expresses his sincere gratitude to Prof. Burda, Zdzisław for the stimulating discussions on matrix theory.